\documentclass[prl,preprint,showpacs,epsf]{revtex4}
\usepackage{graphicx}
\usepackage{color}
\begin{document}

\title{Coulomb shifts and  shape changes in the mass 70 region}

\author{B.S. Nara Singh}
\affiliation{Department of Physics, University of York, Heslington, York YO10
5DD, UK}

\author{A.N.~Steer}
\affiliation{Department of Physics, University of York, Heslington, York YO10
5DD, UK}

\author{D.G.~Jenkins}
\affiliation{Department of Physics, University of York, Heslington, York YO10
5DD, UK}

\author{R.~Wadsworth}
\affiliation{Department of Physics, University of York, Heslington, York YO10
5DD, UK}

\author{M.A.~Bentley}
\affiliation{Department of Physics, University of York, Heslington, York YO10
5DD, UK}

\author{P.J.~Davies}
\affiliation{Department of Physics, University of York, Heslington, York YO10
5DD, UK}

\author{R.~Glover}
\affiliation{Department of Physics, University of York, Heslington, York YO10
5DD, UK}

\author{N.S.~Pattabiraman}
\affiliation{Department of Physics, University of York, Heslington, York YO10
5DD, UK}

\author{C.J.~Lister}
\affiliation{Physics Division, Argonne National Laboratory, 9700 South Cass
Ave., Argonne, IL 60439}

\author{T.~Grahn}
\affiliation{Department of Physics, University of Jyv$\ddot {\rm a}$skyl$\ddot
{\rm a}$,
P.O. Box 35, FIN-40351, Jyv$\ddot {\rm a}$skyl$\ddot {\rm a}$, Finland}

\author{P.T.~Greenlees}
\affiliation{Department of Physics, University of Jyv$\ddot {\rm a}$skyl$\ddot
{\rm a}$,
P.O. Box 35, FIN-40351, Jyv$\ddot {\rm a}$skyl$\ddot {\rm a}$, Finland}

\author{P.~Jones}
\affiliation{Department of Physics, University of Jyv$\ddot {\rm a}$skyl$\ddot
{\rm a}$,
P.O. Box 35, FIN-40351, Jyv$\ddot {\rm a}$skyl$\ddot {\rm a}$, Finland}

\author{R.~Julin}
\affiliation{Department of Physics, University of Jyv$\ddot {\rm a}$skyl$\ddot
{\rm a}$,
P.O. Box 35, FIN-40351, Jyv$\ddot {\rm a}$skyl$\ddot {\rm a}$, Finland}

\author{S.~Juutinen}
\affiliation{Department of Physics, University of Jyv$\ddot {\rm a}$skyl$\ddot
{\rm a}$,
P.O. Box 35, FIN-40351, Jyv$\ddot {\rm a}$skyl$\ddot {\rm a}$, Finland}

\author{M.~Leino}
\affiliation{Department of Physics, University of Jyv$\ddot {\rm a}$skyl$\ddot
{\rm a}$,
P.O. Box 35, FIN-40351, Jyv$\ddot {\rm a}$skyl$\ddot {\rm a}$, Finland}

\author{M.~Nyman}
\affiliation{Department of Physics, University of Jyv$\ddot {\rm a}$skyl$\ddot
{\rm a}$,
P.O. Box 35, FIN-40351, Jyv$\ddot {\rm a}$skyl$\ddot {\rm a}$, Finland}

\author{J.~Pakarinen}
\affiliation{Department of Physics, University of Jyv$\ddot {\rm a}$skyl$\ddot
{\rm a}$,
P.O. Box 35, FIN-40351, Jyv$\ddot {\rm a}$skyl$\ddot {\rm a}$, Finland}

\author{P.~Rahkila}
\affiliation{Department of Physics, University of Jyv$\ddot {\rm a}$skyl$\ddot
{\rm a}$,
P.O. Box 35, FIN-40351, Jyv$\ddot {\rm a}$skyl$\ddot {\rm a}$, Finland}

\author{J. Sar\'en}
\affiliation{Department of Physics, University of Jyv$\ddot {\rm a}$skyl$\ddot
{\rm a}$,
P.O. Box 35, FIN-40351, Jyv$\ddot {\rm a}$skyl$\ddot {\rm a}$, Finland}

\author{C.~Scholey}
\affiliation{Department of Physics, University of Jyv$\ddot {\rm a}$skyl$\ddot
{\rm a}$,
P.O. Box 35, FIN-40351, Jyv$\ddot {\rm a}$skyl$\ddot {\rm a}$, Finland}

\author{J.~Sorri}
\affiliation{Department of Physics, University of Jyv$\ddot {\rm a}$skyl$\ddot
{\rm a}$,
P.O. Box 35, FIN-40351, Jyv$\ddot {\rm a}$skyl$\ddot {\rm a}$, Finland}

\author{J.~Uusitalo}
\affiliation{Department of Physics, University of Jyv$\ddot {\rm a}$skyl$\ddot
{\rm a}$,
P.O. Box 35, FIN-40351, Jyv$\ddot {\rm a}$skyl$\ddot {\rm a}$, Finland}

\author{P.A.~Butler}
\affiliation{Oliver Lodge Laboratory, University of Liverpool, Liverpool L69
7ZE, UK}

\author{M.~Dimmock}
\affiliation{Oliver Lodge Laboratory, University of Liverpool, Liverpool L69
7ZE, UK}

\author{D.T.~Joss}
\affiliation{Oliver Lodge Laboratory, University of Liverpool, Liverpool L69
7ZE, UK}

\author{J.~Thomson}
\affiliation{Oliver Lodge Laboratory, University of Liverpool, Liverpool L69
7ZE, UK}

\author{B.~Cederwall}
\affiliation{Royal Institute of Technology, Roslagstullsbacken 21, S-106 91
Stockholm,
Sweden}

\author{B.~Hadinia}
\affiliation{Royal Institute of Technology, Roslagstullsbacken 21, S-106 91
Stockholm,
Sweden}

\author{M.~Sandzelius}
\affiliation{Royal Institute of Technology, Roslagstullsbacken 21, S-106 91
Stockholm,
Sweden}

\date{\today}
\begin{abstract}
{The technique of recoil beta tagging has been developed which 
allows prompt $\gamma$ decays in nuclei from excited states to be 
correlated with electrons from their subsequent short-lived $\beta$ decay. 
This technique is ideal for studying nuclei very far from 
stability and improves in sensitivity for very short-lived 
decays and for high decay Q-values.  The method has
allowed excited states in $^{78}$Y to be observed for the first time,
as well as an extension in the knowledge of $T~=~1$ states in $^{74}$Rb.
From this new information it has been possible to compare Coulomb
energy differences ({\small CED}) between $T~=~1$ states 
in $^{70}$Br/$^{70}$Se, $^{74}$Rb/$^{74}$Kr and $^{78}$Y/$^{78}$Sr.  
The $A~=~70$ {\small CED} exhibit an anomalous behaviour which is 
inconsistent with all other known {\small CED}. 
This behavior may be accounted for qualitatively in terms of small 
variations in the Coulomb energy arising from shape changes.}

\end{abstract}
\pacs{21.30.-x, 21.10.Sf, 23.20.Lv, 23.40.-s}
\maketitle

The ability of some atomic nuclei to assume competing mean-field shapes
at low excitation energies is a remarkable feature of quantal objects
and is called shape coexistence.  In certain nuclei, a rearrangement of
a few nucleons into different orbitals around the Fermi surface
can result in a substantial change in the energetically favored shape.
One of the classic examples is $^{186}$Pb \cite{and00}, where configurations
resulting in two completely different (prolate and oblate) shapes occur
within 700~keV of the spherical ground state configuration.
An interplay of nuclear shapes is also found in nuclei with mass ($A$) 
around 70 with nearly equal numbers of neutrons ($N$) 
and protons ($Z$), where large shell gaps exist at both oblate {\it and} 
prolate shape for $N~=~Z~=~34$ and 36.
For example, the moments of inertia of the ground state band of $^{68}$Se
suggests an evolution from oblate to prolate shape as a function
of excitation energy \cite{fis00}.  Conversion electron~\cite{bou03} and
Coulomb excitation~\cite{gad05} measurements on $^{72}$Kr also indicate 
shape coexistence.  Such coexisting shapes can lead to long lived isomers,
which could provide bypass routes for the traditional {\it rp}-process 
waiting-points influencing the nucleosynthesis and the 
timescale of X-ray bursts~\cite{schatz, sun}. 
Thus, understanding the interplay of co-existing shapes
provides a sensitive test of our knowledge of nuclear structure 
and has astrophysical significance.

The study of shape coexistence in $N~\sim~Z$ nuclei with $A~\sim~$70
is challenging, as they lie far from stability and are 
difficult to synthesize.  Radioactive beam Coulomb excitation 
is a promising approach for their studies~\cite{hur06}.  
In this Letter, we discuss a technique recently developed 
by us for isolating nuclei in this region through recoil beta 
tagging~\cite{ste06}, and have used it to 
explore Coulomb energy differences ({\small CED}) between 
isospin $T~=~1$ states in odd-odd $N~=~Z$ nuclei ($T_z=(N-Z)/2$=0) 
and their analog states in their 
even-even neighbors.  The {\small CED} is defined as
{\small \it CED}$(J)~=~E_x(J, T=1, T_{z<}) - E_x(J, T=1, T_{z>})$,
where $T_{z>}~=~T_{z<}+1$, $E_x$ is the excitation energy of the states 
of spin $J$ and $T_{z>}~=~(N-Z)/2$ which may take values of 0 or 1 
in this case~\cite{ben06}.  The {\small CED} are driven by effects 
which break charge-symmetry and
charge-independence, the dominant contribution to which is expected to
come from the Coulomb interaction.
They are also exquisitely sensitive to small 
structural changes and, in the present work, reveal 
evidence for variations in shapes in analog states in an 
isospin multiplet.


Over the last decade, the recoil-decay tagging technique~\cite{sim86,pau95} 
({\small RDT}), has become one
of the principal experimental tools for studying nuclei 
at the limits of stability with low production cross sections. 
It employs a recoil separator to separate fusion
residues from primary and scattered beam particles, and fission
products in the case of heavy nuclei.  The residues are subsequently implanted
at the focal plane in a silicon strip detector.  Exotic nuclei close
to the proton drip line are then selected by tagging on
their characteristic $\alpha$-particle or proton emission
following implantation events, and are correlated with $\gamma$ rays 
detected at the target position $\sim$1~$\mu$s earlier,
corresponding to the flight time through the separator.
The possibility of tagging with electrons (or positrons) from
$\beta$-decaying recoils has not been pursued prior to the work
described here.  This is largely due to $\beta$ decay being a 
three-body process where the neutrino (anti neutrino) removes 
some of the energy.  There is, therefore, no characteristic 
$\beta$-particle energy to employ as a tag.  Instead, there 
is a Fermi-Kurie distribution of energies
which, in general, overlaps with the distribution from other
reaction channels.  Fermi super-allowed
$\beta$-emitters constitute a special case with 
exceptionally high $\beta^+$ end-point
energies ($Q(EC)\sim$10~MeV) and short half lives ($<$100 ms).
This Letter reports on the first use of their properties
as a means of channel selection to identify excited states in the
odd-odd $N~=~Z$ nuclei, $^{74}$Rb and $^{78}$Y.


The K130 cyclotron at University of Jyv$\ddot {\rm a}$skyl$\ddot {\rm a}$
accelerated beams of $^{36}$Ar to 103~MeV and $^{40}$Ca to 118
and 121~MeV.  These beams of 4 and 5 particle-nA were incident on
$\sim$1~mg/cm$^2$ $^{nat}$Ca targets for periods of 90 and 210 h
producing $^{74}$Rb and $^{78}$Y, respectively, via the {\it pn}
fusion evaporation channels.  Prompt $\gamma$ rays were recorded
with the {\small JUROGAM} array of 43 Compton suppressed
high-purity germanium detectors
with a total efficiency of 4\% at 1.3~MeV.  Fusion evaporation residues
were separated from the primary beam in the {\small RITU} gas-filled
recoil separator and were implanted in a 700-$\mu$m-thick double-sided
silicon strip detector ({\small DSSSD}) in the {\small GREAT} focal
plane spectrometer~\cite{great}.  Situated behind the {\small DSSSD}
was a planar germanium detector with a thickness of 15~mm.  The combination
of these two detectors served as a $\Delta E-E$ telescope
for recording positrons.  In each case, the {\it pn}
evaporation channel involving Fermi super-allowed $\beta$ decay,
was selected by demanding the detection of a high energy positron,
in a short  ($\sim$100 ms) time coincidence with the implanted recoil.
By correlating with in-beam $\gamma$ rays, recorded in {\small JUROGAM}
$\sim$~1~$\mu$s earlier, it was possible to study the
decay of excited states in $^{74}$Rb and {\it for the first time} in $^{78}$Y.
We refer to this method as recoil beta tagging ({\small RBT}).


Previous in-beam studies of $^{74}$Rb have been carried out using
charged particle and neutron detection for channel 
selection~\cite{rud96,ole03}.  This nucleus has a ground state that 
$\beta$ decays with a half life of 65~ms and an end point energy 
of 9.4~MeV and therefore serves as an excellent 
test case for the {\small RBT} technique.
From the $^{36}$Ar+$^{40}$Ca reaction data, transitions in $^{74}$Rb 
were identified by correlating them with residues implanted
at the focal plane, which were succeeded by the detection of a
positron within $\sim$~100~ms.  Such positrons had to
record an energy loss in the {\small DSSSD} and deposit between
3 and 10~MeV in the planar germanium detector. 
The strong suppression of contaminating channels by the short 
correlation time meant that it was possible to set such a low 
limit (3~MeV) on the positron energy. In this manner, all the 
$\gamma$ rays observed in Refs.~\cite{rud96,ole03} were confirmed, 
and, in particular, the 575 and 478~keV $\gamma$ rays establish the 
energy of the 4$^+$, 2$^+$ states to be 1053
and 478~keV, respectively.  In a recent publication, we focus on the
technique in detail and explore strategies for optimising the
efficiency and cleanliness of the correlations~\cite{ste06}.
The use of a $^{36}$Ar-induced reaction with a beam energy around
the Coulomb barrier, resulted in greater feeding of low-lying
non-yrast states.  This led to the extension of the $T~=~1$ 
ground state band to $J^{\pi}$~=~6$^{+}$ at 1837~keV.  A recently 
published parallel work using more conventional techniques 
confirmed this assignment and found a candidate $J^{\pi}$~=~8$^{+}$  
member of this $T~=~1$ analog sequence \cite{Fis06}. 
The present work has also located a number of additional 
$T~=~0$ states but discussion of these lies beyond 
the scope of this Letter. 

\begin{figure}[htb]
\includegraphics[scale=0.35]{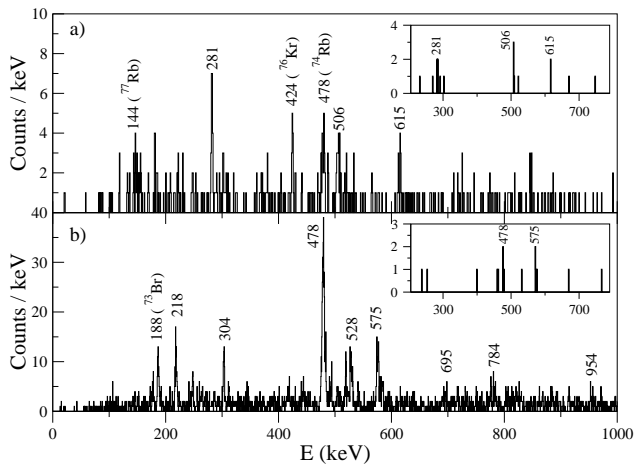}
\hfill \caption{Recoil-$\beta$ tagged $\gamma$ ray spectra
for a) $^{78}$Y and b) $^{74}$Rb~\cite{ste06}. 
The insets show a sum of gates on the 281, 506 and
615~keV, and a gate on the 784~keV transitions, 
in their respective recoil-$\beta$ tagged $E_{\gamma}$-$E_{\gamma}$
matrices.  Details of the time correlations and positron 
energy limits are given in the text.}
\label{fig:fig1}
\end{figure}


Prior to this work the knowledge on $^{78}$Y 
was limited to the $T~=~1$, J$^\pi$~=~$0^+$ ground state 
with its characteristic $T_{1/2}$~=~55(12)~ms superallowed 
$\beta$-decay with an endpoint energy of 9.4~MeV, 
and a 5$^+$ isomer with $T_{1/2}$~=~5.8(6)~s 
\cite{uus98, lon96}.  The isomer receives most of the population 
in the current study using the $^{40}$Ca+$^{40}$Ca reaction; 
implant-decay correlations for high energy 
positrons suggest that as much as 90~$\%$ feeds the isomer.  
Although the isomer 
$\beta$ decays with a high endpoint energy, the half life is  
too long to correctly correlate the decay with the parent 
implant and its associated prompt $\gamma$ rays, since the implantation 
rate per pixel in the {\small DSSSD} was $\sim$~1/s.  However, the 
superallowed decay of the ground state did allow effective 
correlations, and identification of prompt $\gamma$ rays, as was 
achieved for $^{74}$Rb.  The data are shown in Fig.~\ref{fig:fig1}a. 
The lower statistics achieved for the $^{78}$Y study can be 
mainly attributed to population of the isomeric state, as the 
production cross sections for $^{40}$Ca($^{36}$Ar,pn)$^{74}$Rb 
and $^{40}$Ca($^{40}$Ca,pn)$^{78}$Y are expected to be quite similar.

\begin{figure}[htb]
\includegraphics[scale=0.34]{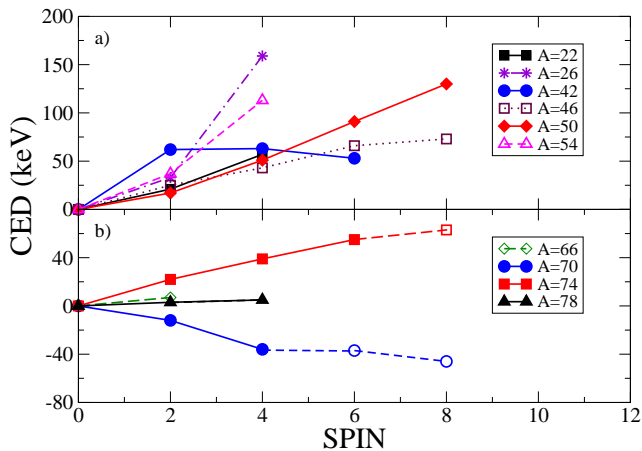}
\hfill \caption{(Color online) {\small CED} between $T_z$~=~(0, 1)
pairs as a function of spin for the
cases: a)~$A~=~$22, 26, 42, 46, 50 and 54 and
b)~$A~=~$ 66, 70, 74 and 78. 
Data for (a) and $A~=~66$ were taken from Refs.~\cite{ben06} 
and ~\cite{Grzy}, respectively .  
Open symbols and dashed lines for (b) represent 
tentatively assigned levels in the 
$N~=~Z$ system considered.}
\label{fig:CEDcombo}
\end{figure}

The low cross section for population of states built 
on the $^{78}$Y ground state made the  breakthrough in 
channel selection using {\small RBT} more apparent 
than in the $^{74}$Rb study.  In this case, it was demanded that the 
$\beta$ particle energy was between 4.5 to 10~MeV within a 
correlation time of 150~ms.  After eliminating 
known transitions from interfering contaminants, 
including $^{74}$Rb produced via the $\alpha pn$ channel, 
three new $\gamma$ lines were identified as belonging to the 
short lived, high endpoint reaction product, $^{78}$Y.  
They were strong enough for it to be 
established that they are in prompt coincidence 
(inset to Fig ~\ref{fig:fig1}a). 
The intensity of the 281~keV $\gamma$ ray is consistent with it being the
strongest transition and hence it is most likely to decay
to the ground state.  This could be shown to be associated
with positrons decaying with a halflife of 47(5)~ms, in good agreement 
with the known $^{78}$Y ground state decay.  The angular distributions
of the two stronger lines were consistent with quadrupole multipolarity
although with large uncertainties due to poor statistics.  
It is also a common practice in studies of isobaric analog nuclei to assume that 
the analogue states have a similar structure at a given spin~\cite{ben06_2}.
Thus, we tentatively assign the 506 and 281~keV transitions as 
the analogs of the 504 and 278~keV $\gamma$ rays 
corresponding to the 4$^{+}$$\rightarrow$2$^{+}$$\rightarrow$0$^{+}$ 
cascade in the $T~=~1$ ground state band in $^{78}$Sr. 
Unfortunately, we are unable to determine the multipolarity of the 
615~keV $\gamma$ ray.  Moreover, we note that if this is assumed to be 
the $T~=~1$, 6$^{+}$$\rightarrow$4$^{+}$ transition, then it would result 
in a large negative {\small CED} of -92~keV compared to the small positive
values for the 2$^+$ and 4$^+$ states (see Fig.~\ref{fig:CEDcombo}).  
Whilst the systematics suggest that such an abrupt change 
is not impossible, it could also be that the 615~keV $\gamma$ decay 
does not originate from the 6$^+$ member of $T~=~1$ sequence.

The new {\small CED} data on $A~=~{74}$ and 78 nuclei can now be compared to 
the published $A~=~70$ trend \cite{DeAn01, jen02}, as shown 
in Fig.~\ref{fig:CEDcombo}b.  They show a remarkable contrast.  
The {\small CED} falls for $A~=~70$, rises 
smoothly for $A~=~74$ and is near-zero (at least up to spin 4) for $A~=~78$. The $A~=~70$ trend was 
previously attributed \cite{DeAn01} to the Thomas-Ehrman effect; 
the loosely bound proton in $^{70}$Br was anticipated to have an 
unusually extended radial wavefunction and thus have a lower 
Coulomb energy than the equivalent state in $^{70}$Se.  In the 
light of the new data, this cannot be the complete explanation, 
as all the three systems have similar differences in binding energy 
between the $T_z~=~0$ and $T_z~=~1$ nuclei, and so should all exhibit the 
same trend.  Moreover, these states are all well bound so are unlikely to 
have significantly extended wavefunctions.

The trend in {\small CED} across the $sd$ and $fp$ shell has been investigated 
in considerable detail in recent years.  New data have been 
obtained and interpreted using large scale shell model 
calculations~\cite{len99,ole02,ben06}.  As shown in 
Fig.~\ref{fig:CEDcombo}a, the 
{\small CED} have a positive trend in the $sd$ and $fp$ shells.
The microscopic explanation for this ubiquitous trend lies 
in the destruction of pairing correlations by angular momentum,
i.e. Coriolis anti-pairing.  For perfect charge independence, 
this destruction should be exactly the same in $T_z$~=~0 and $T_z$~=~1 
nuclei of same mass; the generation of angular momentum reduces the occupation 
of exactly time reversed orbits and the overlap of wavefunctions 
is diminished.  For proton pairs, this lowers the Coulomb energy. 
Depending on how many proton-proton pairs are being destroyed, 
there can be a small difference in the Coulomb energy change between 
analog nuclei.  For the $N~=~Z$, $T_z$~=~0 nuclei, it is expected  
that neutron-proton $T~=~1$ pairing correlations are important. For the 
$N~=~Z+2$ nuclei with $T_z$~=~1 only $nn$ and $pp$ pairs are expected to 
play a significant role~\cite{len99, ole02, warner}.  Thus, there are always 
more proton-proton pairs in the $N~=~Z+2$ nuclei, and consequently a larger 
reduction in Coulomb energy with spin.  In a large shell model 
space, or a single $j$-shell with many pairs of particles and 
a high level density, the {\small CED} would rise smoothly with spin. 
Empirically, this effect is 10-15~keV per unit of 
angular momentum.  However, in the restricted spaces  
for intermediate mass nuclei with several orbitals 
of differing $j$, the effect can become irregular depending on the 
microscopic construction of the pairs, particularly the angular 
momentum of the underlying single particle states.

The $A~=~70$ case, with its unique negative {\small CED} needs a new 
explanation.  The interpretation of the {\small CED} behavior in the 
$fp$-shell assumes that the nuclear shapes are fixed and 
that $T~=~1$ $np$-pairing is important only in $N~=~Z$ nuclei. 
If either of these considerations are not valid, then the {\small CED} can
assume a different behavior.  Experiments on $^{68}$Se and $^{72}$Kr 
show evidence for the presence of an oblate shape at low excitation 
energy~\cite{fis00,bou03,gad05}.  In
the neighbouring nuclei, $^{70}$Se~\cite{myl89, hee86} and
$^{74}$Kr~\cite{Rudo1, bend06}, shape changes have also been suggested
to play an important role.  With the assumption of charge 
independence, the shape coexistence {\it must be} exactly the 
same for the $T~=~1$ states in the odd-odd $N~=~Z$ nuclei, $^{70}$Br and
$^{74}$Rb.  To lowest order, the spectra should then be identical. 
However, this does not imply the {\small CED} will
be zero, as the Coulomb monopole cancellation obtained by aligning 
the ground state energies, only remains exact if the shapes
remain frozen.  Any evolution of shape with spin (including stretching 
or changes in shape) will perturb the {\small CED}. 
Specifically, big increases in deformation with spin 
lead to negative {\small CED}.  Thus, negative {\small CED} provide new and 
sensitive information on shape evolution.

We have investigated the effects of shape change on the {\small CED} 
using a deformed liquid drop model~\cite{Lars} and calculated 
the effects for $A~=~70$.  Shape changes in $^{70}$Se are clear 
experimentally from the very irregular yrast
line and lifetime measurements, which indicated a strong 
reduction in $B(E2)$ transition strengths near $J~=~4$~\cite{hee86}. 
A recent Coulomb excitation measurement
favors a prolate shape for the 2$^{+}$ state consistent with
$\beta_2$~=~0.25~\cite{hur06} and is in agreement 
with a configuration mixing shell model calculation
which predicts a ground state band with $\beta_2$~=~0.18 
crossed by a more deformed band with
$\beta_2$~=~0.33 near $J~=~6$~\cite{sah87}. 
For such a shape change, the deformed liquid drop
model suggests a $\sim$ 75~keV decrease in {\small CED}, in good 
agreement with that observed in the present work. 
Historically, the shape coexistence in $^{70}$Se has been described as
a competition between an oblate ground state configuration 
and an excited prolate configuration; this interpretation 
being supported by Total Routhian Surface 
calculations~\cite{myl89}. However, for such a shape 
change, i.e. $\beta_2$ from -0.3 (oblate) to 0.35 (prolate), 
the {\small CED} should only decrease by $\sim$~7~keV, which does not
account for the observed trend.  Only a significant change in
elongation can make sufficient change to the Coulomb energy.

The $A~=~74$ {\small CED} reveal a monotonically positive trend.  
This seems to imply that in the $T~=~1$ band the deformation 
up to $J~=~8$ is always large and does not change significantly
(supported by $B(E2)$ data~\cite{Gorg}), 
so the {\small CED} evolution is due to Coriolis anti-pairing 
as in the case of $f_{7/2}$ nuclei. 
Beyond the coexistence region, in the middle of the $fpg$ shell at 
$A~\sim~$80, very large and stable prolate deformation is known 
to be stabilized by a gap in the single particle sequence 
at $N~=~Z~=~38$~\cite{Rudo1}.  The gap is sufficiently large 
that all scattering across the Fermi surface is suppressed 
and the odd-$A$ nuclei 
appear as near rigid rotors \cite{Har97}.  With stable shapes 
and all pairing effects reduced the very small {\small CED} found in 
the $^{78}$Y-$^{78}$Sr pair at low spin are perhaps not surprising. 
However, as the proton backbend, which occurs at $J~=~8$~\cite{rud96},
is approached in $^{78}$Sr one may well expect a change in {\small CED} to appear.
This could explain the {\small CED} of -92~keV discussed earlier if the 615~keV 
transition in $^{78}$Y originates from the decay of the $T~=~1$, 6$^+$ state.

In summary, information on Coulomb energy differences in $T~=~1$ 
multiplets has been extended using recoil beta tagging. 
The {\small CED} derived for $A~=~70$ are quite different 
from the expectations based on our knowledge of the 
behavior in the $fp$-shell.  We suggest that the pronounced 
decrease in {\small CED} as a function of spin is due to subtle differences 
in the Coulomb energy as shapes evolve with spin.  If this is the correct 
explanation, it will be manifested in the $T_z=-1$ member of the isobaric
triplet through a further lowering of the ground state band of $^{70}$Kr,
by an amount equal to that observed between the $^{70}$Br/$^{70}$Se pair,
as the effect is linear with $Z$.  Currently nothing is known about 
the excited states in $^{70}$Kr, but clearly this becomes an important 
measurement.  From a theoretical stand point, these measurements provide a 
definitive test case for state-of-the-art shell model calculations.
Even though the 2$^+$ level assignment in $^{66}$As is tentative, 
it is also interesting to note that for the $A~=~66$ pair 
($^{66}$As/$^{66}$Ge-see Fig.~\ref{fig:CEDcombo}b) the {\small CED} for 
the $T~=~1$, 2$^+$ states is about zero, suggesting a trend similar 
to that for $^{78}$Y.  Clearly, extending these data for the $T~=~1$ 
states in $^{66}$As will be important, since if the {\small CED} remains near zero 
as a function of spin in this case then a different explanation to that given 
above for the mass 78 pair will be required.
During the preparation of this Letter, we learnt about studies 
of $^{82}$Nb and $^{86}$Tc~\cite{garn07}, which promise 
further insight into the issues discussed here.


This work was partly supported by the {\small UK EPSRC} 
and by the {\small INTAG JRA} 
and {\small TNA} within the {\small EURONS} Integrated Infrastructure 
Initiative ({\small R113-CT-2004-506065}) and  by the 
{\small US} Department Of Energy, Office of Nuclear Physics, 
under contract {\small W-ENG}-109-38.

\end{document}